\documentclass[english,letterpaper,aps,prl,twocolumn]{revtex4}
\usepackage{lmodern}
\usepackage[T1]{fontenc}
\usepackage[latin1]{inputenc}
\usepackage{amsmath}
\usepackage{graphicx}

\makeatletter
\@ifundefined{textcolor}{}
{%
 \definecolor{BLACK}{gray}{0}f
 \definecolor{WHITE}{gray}{1}
 \definecolor{RED}{rgb}{1,0,0}
 \definecolor{GREEN}{rgb}{0,1,0}
 \definecolor{BLUE}{rgb}{0,0,1}
 \definecolor{CYAN}{cmyk}{1,0,0,0}
 \definecolor{MAGENTA}{cmyk}{0,1,0,0}
 \definecolor{YELLOW}{cmyk}{0,0,1,0}
}

\usepackage{lmodern}

\makeatletter

\usepackage{epsfig}

\usepackage{dcolumn}

\usepackage{bm}

\usepackage{bbm}

\usepackage{wasysym}

\usepackage{marvosym}

\usepackage{subfloat}
\usepackage{color}
\usepackage{subfigure}

\usepackage{blindtext}

\newlength{\textwidthm}

\usepackage{tikz}

\makeatother

\usepackage{babel}
\usepackage{pgffor}
\usepackage{pdfpages}
\begin{document}
\title{Strongly Interacting Phases in Twisted Bilayer Graphene at the Magic
Angle}
\author{Khagendra Adhikari$^{1}$, Kangjun Seo$^{2}$, K. S. D. Beach$^{3}$
and Bruno Uchoa$^{1}$}
\email{uchoa@ou.edu}

\affiliation{$^{1}$Department of Physics and Astronomy, University of Oklahoma,
Norman, OK 73069, USA}
\affiliation{$^{2}$School of Electrical and Computer Engineering, University of Oklahoma, Tulsa, OK 74135, USA}
\affiliation{$^{3}$Department of Physics and Astronomy, The University of Mississippi,
University, MS 38677, USA}

\date{\today}

\date{\today}
\begin{abstract}
Twisted bilayer graphene near the magic angle is known to have a
cascade of insulating phases at integer filling factors of the low-energy bands.
In this Letter we address the nature of these phases
through an unrestricted, large-scale Hartree-Fock calculation on
the lattice that self-consistently accounts for all electronic bands. Using numerically
unbiased methods, we show that Coulomb interactions 
produce ferromagnetic insulating states at integer
fillings $\nu\in[-3,3]$ with maximal spin polarization $M_{\text{FM}}=4-|\nu|$.
We find that the $\nu=0$ state is a pure ferromagnet, whereas all other insulating states are spin-valley polarized.  
At odd filling factors
 $|\nu|=1,3$ those states have a quantum anomalous Hall effect with Chern number $\mathcal{C}=1$.  Except for the $\nu=0,-2$ states,
all other integer fillings
have insulating phases with additional sublattice symmetry breaking
and antiferromagnetism in the remote bands.
We map the metal-insulator transitions of 
 these phases as a function of the effective
dielectric constant. Our results establish the importance of large-scale lattice calculations to faithfully determine the ground states of TBG at integer fillings.
\end{abstract}
\maketitle
{\it Introduction}.---When two graphene sheets are twisted by a
small angle, dubbed the ``magic angle,'' their low energy bands reconstruct
into flat bands \cite{Santos,Bistrizer}. A cascade of insulating
phases has been recently observed in twisted bilayer graphene (TBG)
near the magic angle at integer filling fractions of the flat bands
\cite{Cao1,Lu,Zondiner,Morissette,Park}, some of which are in proximity
to low-temperature superconducting phases \cite{Cao0,Yankowitz,Stepanov,Jaoui,Saito}.
Magic angle TBGs are considered strongly correlated electron systems,
with evidence of a strange metal phase at finite temperature \cite{Polshyn,Cao3}
and of low-energy electronic collective modes that are strongly coupled
to the charge carriers \cite{Saito-2}. Because of the quantum geometry
of the flat bands and the degeneracy between spin, valley, and sublattice
degrees of freedom in graphene \cite{Kotov}, interactions in combination
with magnetic field, strain and substrate effects can drive various
strongly interacting phases. Recent experiments observed evidence
of quantum anomalous Hall (AQH) states \cite{Spanton,Sharpe,Serlin,Wu},
charge order \cite{Jiang}, ferromagnetism \cite{Sharpe}, and Kekule
intervalley coherent states \cite{Nuckolls,Wagner,Kwan} in the presence
of strain. 

The nature of intrinsic correlated insulating phases at zero field,
in the absence of strain or substrate effects, is not clear. 
For instance, the
insulating state at half-filling ($\nu=0$) is unknown. Quantum Monte Carlo results in effective models predicted 
the possibility of valley quantum Hall \cite{Liao, Hofmann}, intervalley coherent  (IVC) states and a valence bond solid \cite{Liao}. 
Hartree-Fock (HF) calculations found an antiferromagnet \cite{Gonzalez}, a valley Chern insulator \cite{Liu} and an IVC state \cite{Bultinick}.
Very little numerical results exist about the $|\nu|=1$ states.
 A HF calculation found a Chern insulator with IVC states \cite{Wagner}. At $\nu=-2$, HF \cite{Gonzalez} and exact diagonalization \cite{Potasz}
results encountered a pure spin polarized ferromagnet, whereas other HF results proposed a pure valley polarized state \cite{M Xie} and also an IVC state \cite{Wagner}. 
A growing consensus between HF and exact diagonalization at  $|\nu|=3$ indicates that this state may be a spin-valley polarized Chern insulator \cite{F Xie, Potasz, F Xie 2}. 
The determination of these phases could shed light in the nature of the superconducting states in TBG and remains an outstanding open question.

At the magic angle $\theta\approx1.1^{\circ}$, the Moir\'{e} unit cell
with size $L\sim a/[2\sin(\theta/2)]\approx12$ nm, where $a$ is
the graphene lattice constant, has $\sim10^{4}$ lattice sites, making
\emph{ab initio} methods prohibitively expensive. Previous numerical
studies based on Hartree-Fock \cite{Bultinick,M Xie,F Xie,Liu,Cea,Liu-1},
exact diagonalization \cite{Potasz,F Xie 2,Reppelin}, quantum Monte
Carlo \cite{Liao,Hofmann}, and density matrix renormalization group
\cite{Soejima,Kang} relied on effective low-energy models in the
continuum approximation, or else in effective lattice models. The latter were derived
either with the extraction of Wannier orbitals from the low-energy
flat bands \cite{Kang-1,Koshino,Kang3,Seo,Breio, Andrews}, phenomenologically \cite{Khalifa}, or else with the introduction
of an energy cutoff in the remote bands \cite{Gonzalez}. 

Unlike in graphene 
monolayer, where continuum models are justified by the separation of the valleys through kinematic constraints \cite{Kotov}, in TBG the valleys  are not good 
quantum numbers even in the absence of backscattering \cite{note2, Ramires}.  Besides, at length
scales shorter than the size of the Moir\'{e} unit cell $L$, Coulomb
interactions can be one order of magnitude larger than the energy
gap separating the low-energy  bands from the remote ones. A more conclusive
determination of the ground state needs to account for virtual processes
that connect the two sets of bands \cite{note3} and may require treating the remote
bands on an equal footing with the flat ones through an unbiased calculation on the lattice.

In this Letter we perform an unrestricted, self-consistent,  large-scale
HF calculation on the lattice with Coulomb interactions screened
only by metallic gates. The calculation fully accounts for all
remote bands in the emergence of many-body states. We find that the
insulating ground states of integer fillings $\nu\in[-3,3]$ are ferromagnetic,
with a magnetic moment $M_{\text{FM}}=4-|\nu|$ due to the
spin polarization of the flat bands. We find that the $\nu=0$ state is a pure spin ferromagnet. 
Valley polarization (VP) is found at all integer $\nu$ away from $\nu=0$.
Except for $\nu=0,-2$ we identified antiferromagnetism
(AF) in the remote bands, which appears with sublattice polarization
(SLP). 
Our results also indicate that odd filling factors $|\nu|=1,3$ have
quantum anomalous Hall (QAH) states with Chern number $\mathcal{C}$=1,
whereas even filling factors $|\nu|=2$ have zero net Chern number. At $\nu<0$ ($\nu >0$), all flat bands with majority (minority) spin are topological, whereas the ones with minority (majority) spin are trivial. All bands are topologically trivial in the $\nu=0$ state.

\begin{figure}
\includegraphics[width=1\columnwidth]{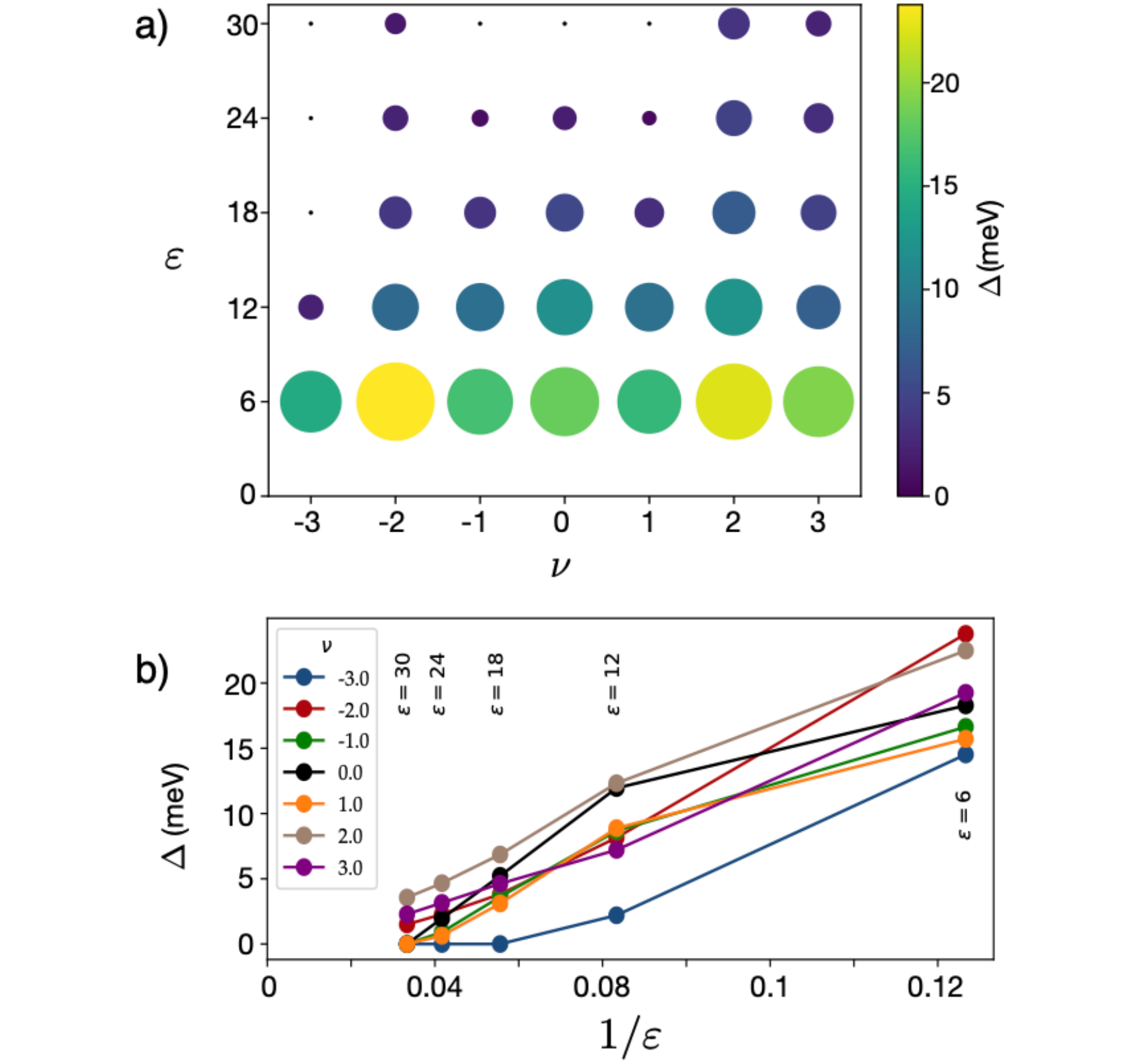}

\caption{{\small{}a) Phase diagram of strongly interacting phases in magic
angle TBG at integer fillings $\nu\in[-3,3]$ of the flat bands for
different values of the effective dielectric constant $\varepsilon$.
Size of the circles and color indicate the amplitude of the many-body
gap $\Delta$. Small dots are metallic states (zero gap), indicating
the presence of a metal-insulator transition for a critical value
of $\varepsilon$. b) Scaling of the many-body gap $\Delta$ vs $1/\varepsilon$
for different integer fillings. }}
 \label{FIG:phase_diagram}
\end{figure}

Those insulating states have a metal-insulator transition (MIT) at
critical values of the effective dielectric constant $\varepsilon$,
as shown in Fig.~\ref{FIG:phase_diagram}. The amplitudes of the many-body gaps for $|\nu|=3$
are strongly particle-hole \emph{asymmetric} in the moderate- and small-gap
 regimes, in agreement with experiments \cite{Zondiner,Morissette} 
 and in contrast with continuum models, which are particle-hole symmetric \cite{note4}.
We conjecture that most experimental samples are in the vicinity of
the $\nu=0$ state MIT. In this regime, we find that the $\nu=\pm2,3$
states have prominent gaps of a few meV, whereas the other filling
factors $\nu=-3,\pm1$ are either metallic or have very small many-body
gaps, also in agreement with experiment. We suggest that these MIT transitions can be experimentally observed.

{\it Hamiltonian}.---We begin with the real space lattice Hamiltonian
of TBG, $\mathcal{H}=\mathcal{H}_{0}+\mathcal{H}_{\text{C}},$ where
\begin{equation}
\mathcal{H}_{0}=\sum_{\alpha\beta}\sum_{ij}h_{\alpha\beta}(\mathbf{R}_{i}-\mathbf{R}_{j})d_{\alpha,i,\sigma}^{\dagger}d_{\beta,j,\sigma}\label{eq:H0}
\end{equation}
is the hopping term, with $d_{\alpha,i,\sigma}$ the annihilation
operator of an electron with spin $\sigma=\uparrow,\downarrow$ on
sublattice $\alpha$ of unit cell $i$, and $h_{\alpha\beta}(\mathbf{R}_{i}-\mathbf{R}_{j})$
is the corresponding tight-binding matrix element between sublattices
$\alpha$ and $\beta$ located in unit cells centered at $\mathbf{R}_{i}$
and $\mathbf{R}_{j}$ respectively. The second term describes a screened
Coulomb interaction between any two lattice sites, 
\begin{equation}
\mathcal{H}_{\text{C}}=\frac{1}{2}\sum_{\alpha\beta}\sum_{ij}\hat{n}_{\alpha,i}V_{\alpha\beta}(\mathbf{R}_{i}-\mathbf{R}_{j})\hat{n}_{\beta,j},\label{eq:Hc}
\end{equation}
where $\hat{n}_{\alpha,i}=\sum_{\sigma}d_{\alpha,i,\sigma}^{\dagger}d_{\alpha,i,\sigma}$
is the local density operator and $V_{\alpha\beta}(\mathbf{R})=e^{2}/(\varepsilon\xi)\sum_{m=-\infty}^{\infty}(-1)^{m}[(|\boldsymbol{\tau}_{\alpha}-\boldsymbol{\mathbf{\tau}}_{\beta}+\mathbf{R}|/\xi)^{2}+m^{2}]^{-\frac{1}{2}}$
is the screened form of the interaction in the presence of symmetric
gates located at the top and bottom of the TBG heterostructure
\cite{Throckmorton,Bernevig2}, with $e$ the electron charge and
$\boldsymbol{\tau}_{\alpha}$ the position of a site in sublattice
$\alpha$ measured from the center of its unit cell. $\xi\approx10$\,nm
is the distance between the plates of the metallic gates in most experiments
\cite{Stepanov,Saito}, and $\varepsilon$ is the effective dielectric
constant, which we treat as a free adjustable parameter. The value
of the dielectric constant of graphene encapsulated in hexagonal boron
nitride (hBN) is $\varepsilon\approx6$, although polarization effects
due to remote bands in TBG could effectively make it several times
larger \cite{Potasz,Gonzalez}. We regularize the on-site Coulomb
interaction by choosing the onsite Hubbard term $V_{\alpha\alpha}(\mathbf{R}=0)=12.4/\varepsilon\,$eV
of single layer graphene \cite{Wehling} and then smoothly interpolating
it with Eq.~\eqref{eq:Hc} through $V_{\alpha\beta}(\mathbf{R})\approx1.438/[\varepsilon(0.116+|\boldsymbol{\tau}_{\alpha}-\boldsymbol{\mathbf{\tau}}_{\beta}+\mathbf{R}|)]$
eV \cite{Radamaker}. Although this choice is not unique, the results
do not depend on the details of the regularization. 

Strain effects can favor ground states that break the translational
symmetry of the Moir\'{e} pattern \cite{Nuckolls,Wagner,Kwan}. To the
best of our knowledge, no broken translation symmetry has been observed
to date in TBG without strain \cite{Xie,Kerelsky}. We assume the
absence of strain effects and rewrite the Hamiltonian in momentum
representation, 
\begin{equation}
\mathcal{H}=\sum_{\alpha\beta}\sum_{\mathbf{k},\sigma}h_{\alpha\beta}(\mathbf{k})d_{\alpha,\mathbf{k},\sigma}^{\dagger}d_{\beta,\mathbf{k},\sigma}+\mathcal{H}_{\text{C}},\label{eq:H0-1}
\end{equation}
where
\begin{equation}
\mathcal{H}_{\text{C}}=\frac{1}{2}\sum_{\alpha\beta}\sum_{\mathbf{q}}\hat{n}_{\alpha}(\mathbf{q})V_{\alpha\beta}(\mathbf{q})\hat{n}_{\beta}(-\mathbf{q}),\label{eq:HC2}
\end{equation}
with $V_{\alpha\beta}(\mathbf{q})$ being the Fourier transform of the Coulomb
interaction in Eq.~\eqref{eq:Hc}. 

\begin{figure*}
\includegraphics[width=1\textwidth]{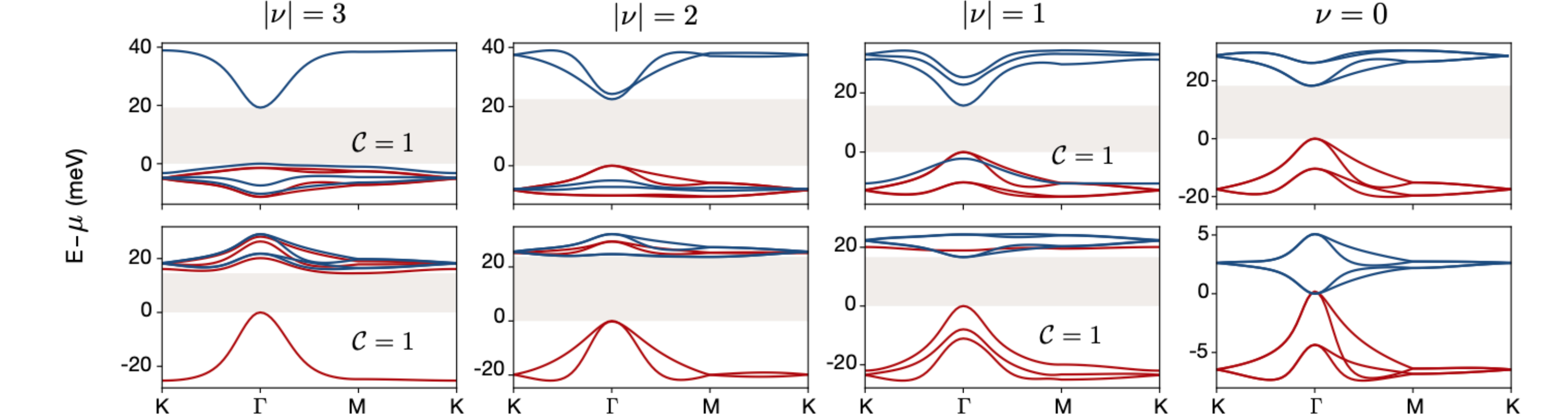}

\caption{{\small{}Band structure of the low energy spin polarized bands for
integer filling factors. Top row from left to right: $\nu=3,2,1,0$
for $\varepsilon=6$. Bottom row (left to right): $\nu=-3,-2,-1$
for $\varepsilon=6$ and $\nu=0$ for $\varepsilon=30$ (metallic
state). The gray shadow indicates the size of the many-body gap. Red (blue) lines represent spin $\sigma$ ($-\sigma$).
For $\nu<0$ ($\nu>0$) all red (blue) bands are topological, while blue (red) bands are trivial.
Odd and even filling factors have a net Chern number $\mathcal{C}=1$
and zero, respectively. 
At $\varepsilon=6$ all integer filling factors have insulating states. 
The insulating state at  $\nu=0$
is ferromagnetic, with zero valley polarization. All non-zero filling
states are spin and valley polarized. The $\nu=0$ state has a MIT
in the range $24<\varepsilon<30$ (see text). }}
 \label{FIG:band_structure}
\end{figure*}

Near the magic angle $\theta\approx1.085^\circ$ the unit 
cell of the commensurate structure has $N_{b}=11164$ lattice sites, and $\alpha=1,\ldots,N_{b}$.
The tight-binding hopping parameters $t(\mathbf{r})$ are calculated from the standard
Slater-Koster functions parametrized by {\it ab initio} calculations \cite{Note,Nam}.
We define $h_{\alpha\beta}(\mathbf{R}_{i}-\mathbf{R}_{j})\equiv t(\mathbf{r}_{\alpha}-\mathbf{r}_{\beta})-\mu\delta_{\alpha\beta}\delta_{ij}$,
with $\mathbf{r}_{\alpha}=\tau_{\alpha}+\mathbf{R}_{i}$ the position
of the lattice sites and $\mu$ the chemical potential. We account
for lattice corrugation effects by allowing the layers to relax in
the $z$ direction. The relaxed tight-binding band structure of TBG
has eight low energy bands with a band width of $\sim 6$\,meV separated
from the remote bands by an energy gap of $20$ meV \cite{Koshino,Nam,Uchida,Note2}.
At the Hartree level, charging effects are expected to significantly
reconstruct the low energy bands \cite{Guinea}. 

{\it Hartree-Fock calculation}.---We proceed to identify the many-body
ground states of the problem through an unrestricted, self-consistent
HF calculation that accounts for \emph{all} the electronic bands. A HF decomposition of the Hamiltonian reduces the problem to one of diagonalizing the effective Hamiltonian 
\begin{equation}
\mathcal{H}_{\text{HF}}=\sum_{\mathbf{k},\sigma}\sum_{\alpha\beta}\bar{h}_{\alpha\beta}(\mathbf{k},\sigma)d_{\alpha,\mathbf{k},\sigma}^{\dagger}d_{\beta,\mathbf{k},\sigma}.\label{eq:H_HF}
\end{equation}
Here $\bar{h}_{\alpha\beta}(\mathbf{k},\sigma)=h_{\alpha\beta}(\mathbf{k})+h_{\alpha\beta}^{\text{H}}(\mathbf{k},\sigma)+h_{\alpha\beta}^{\text{F}}(\mathbf{k},\sigma)$
is the renormalized matrix elements due to both Hartree, 
\begin{equation}
h_{\alpha\beta}^{\text{H}}(\mathbf{k},\sigma)=\delta_{\alpha\beta}\sum_{\gamma,\mathbf{k}^{\prime},\sigma^{\prime}}V_{\beta\gamma}(0)\rho_{\gamma\gamma}(\mathbf{k}^{\prime},\sigma^{\prime}),\label{eq:H_H}
\end{equation}
and Fock contributions, 
\begin{equation}
h_{\alpha\beta}^{\text{F}}(\mathbf{k},\sigma)=-\sum_{\mathbf{k}^{\prime}}V_{\alpha\beta}(\mathbf{k}-\mathbf{k}^{\prime})\rho_{\alpha\beta}(\mathbf{k}^{\prime},\sigma).\label{eq:H_F}
\end{equation}
Each contribution can be cast in terms of the $N_{b}\times N_{b}$
zero-temperature density matrix for a given momentum and spin,
\begin{equation}
\rho_{\alpha\beta}(\mathbf{k},\sigma)=\sum_{n}^{\text{occupied}}\phi_{\alpha,\mathbf{k}}^{(n)}(\sigma)\phi_{\beta,\mathbf{k}}^{(n)*}(\sigma),\label{eq:rho}
\end{equation}
which is defined as a sum over all occupied bands, with $n=1,\ldots,N_{b}$.
The $N_{b}$-component spinors $\phi_{\alpha,\mathbf{k}}^{(n)}(\sigma)$
describe the exact eigenvectors of Eq.~\eqref{eq:H_HF}, which need to
be calculated self-consistently while enforcing a fixed total number of
particles per unit cell. 

The ground state at the HF level is found by self-consistently calculating
the real space density matrix, $\hat{\rho}_{\alpha\beta}(\mathbf{R},\sigma)$,
from a $9\times9$ mesh in momentum space including the $\Gamma$
point. Each self-consistent loop renormalizes the chemical potential
of the previous step to ensure conservation of the total number of
particles. All densities are measured away from the neutrality point.
We subtract the density matrix from a reference density matrix corresponding
to a uniform background of charge. The initial density matrix is 
extracted from the relaxed non-interacting tight binding Hamiltonian
and self-consistently iterated until convergence. We implement a strong
convergence criteria for the density matrix, where the root mean square
of the difference in density matrices with the previous iteration
is less than the tolerance, $\delta\hat{\rho}_{\alpha\beta}(0,\sigma)<10^{-8}$.
In order to probe for many-body instabilities, we seed the initial
density matrix with random noise, allowing for all symmetries (with
the exception of translational symmetry) to be broken. After convergence,
we then perform numerical measurements with the resulting density
matrix to identify the ground states at different filling factors.
This approach provides an unbiased method to probe for many-body instabilities
at the HF level. 

{\it Ferromagnetic ground states}.---In Fig.~\ref{FIG:band_structure} we show the reconstructed
low energy bands of magic angle TBG for integer filling factors $\nu\in[-3,3]$
for $\varepsilon=6$. At this value of $\varepsilon$ we find that
all integer fillings correspond to insulating states. Interactions
increase the bandwidth of the low-energy flat bands by a factor of
$\sim3$, while splitting their spin degeneracy. The lines in red
(blue) are spin $\sigma$ ($-\sigma$) bands, which are fully spin
polarized, with large many-body gaps $\Delta$ (see Fig.~\ref{FIG:phase_diagram}a). For
$\nu=-3$ the ground state is a ferromagnetic insulator with a single
occupied flat band with spin $\sigma$, as depicted in Fig.~\ref{FIG:band_structure}.
At higher non-positive integer fillings $\nu\leq0$ 
the spin polarization $M_{\text{FM}}$ increases in integer increments
and is maximal at half filling ($\nu=0$), where the ground state
has four occupied mini-bands with the same spin $\sigma$. At
positive integer fillings ($\nu>0$) additional mini-bands with
spin $-\sigma$ are occupied (see upper row of Fig.~\ref{FIG:band_structure}). We find that the the
system has maximum total spin polarization $M_{\text{FM}}=4-|\nu|$
for $\nu\in[-3,3]$. 

This behavior is observed in all insulating states
at higher values of $\varepsilon$. For $\varepsilon=12, 18, 24$ and $30$
the many-body gaps progressively decrease while the low-energy bands
become flatter, suggesting a series of filling-factor-dependent MITs
\cite{Note2-1}. For instance, at $\varepsilon=18$ the $\nu=-3$
state is metallic whereas the $\nu=0,\pm1$ states have a MITs in the range  $24<\varepsilon<30$
(see Fig. \ref{FIG:phase_diagram}). The $\nu=\pm2,3$ states are more robust, persisting at much larger values of $\varepsilon$. The scaling of $\Delta$ with $1/\varepsilon$ for all
integer fillings is shown in Fig.~\ref{FIG:phase_diagram}b. 

\begin{figure}[t]
\vspace{0.0cm}
\includegraphics[width=1\columnwidth]{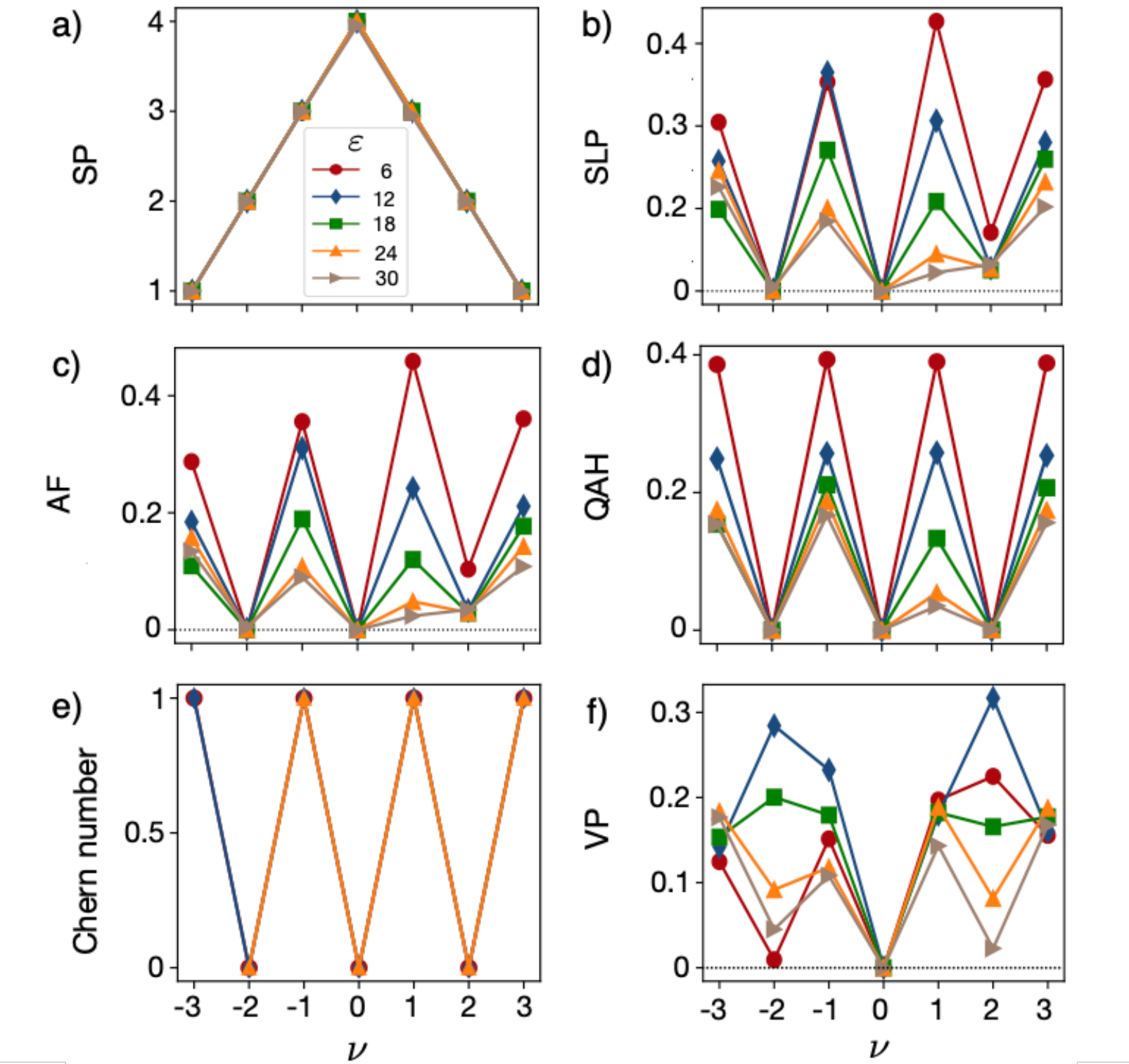}

\caption{{\small{}Magic angle TBG instabilities at $\varepsilon=6$
(red dots), 12 (blue diamonds), 18 (green squares), 24 (orange triangles) and 30 (gray triangles) and for integer
 $\nu\in[-3,3]$: a) Spin polarization, b) sublattice
 polarization, c) antiferromagnetism, d) quantum anomalous
Hall, e) total Chern number 
and f) valley polarization. 
The $\nu=0$ state has zero valley and sublattice polarization (see text). }}
 \label{FIG:measurement}
\end{figure}

The ferromagnetic order parameter can be defined directly from the
real-space density matrix as $M_{\text{FM}}=\sum_{\alpha}s_{\alpha}$,
with $s_{\alpha}=\hat{\rho}_{\alpha\alpha}(0,\sigma)-\hat{\rho}_{\alpha\alpha}(0,-\sigma)$
the local spin polarization summed over all lattice sites in the Moir\'{e}
unit cell. 
Only the flat bands contribute to the magnetization $M_{\text{FM}}$
and to the ensuing ferromagnetic order. The ferromagnetic states are
concentrated in \emph{AA} site regions, where the charge density of
the flat bands is also concentrated \cite{Note3}. We summarize in
Fig.~\ref{FIG:measurement}a the explicit measurement of $M_{\text{FM}}$ through the density
matrix for different integer fillings. 

With the exception of half and quarter filling ($\nu=0,-2$), we find
SLP in all other insulating states. The order parameter for SLP breaks
the symmetry between sublattices $A$ and $B$ in each monolayer and
is defined as $m_{\text{SLP}}=\sum_{\sigma,\alpha\in A}\hat{\rho}_{\alpha\alpha}(0,\sigma)-\sum_{\sigma,\alpha\in B}\hat{\rho}_{\alpha\alpha}(0,\sigma)$.
In Fig.~\ref{FIG:measurement}b we show the contribution to SLP from both the low-energy and
remote bands. We find that SLP is accompanied by a sublattice modulation
of the ferromagnetism in the low-energy bands, and
by a pattern of \emph{staggered} magnetization in the \emph{remote}
bands, indicating AF order \cite{Note3}. In Fig.~\ref{FIG:measurement}c we display the AF order parameter
$M_{\text{AF}}=\sum_{\alpha\in A}s_{\alpha}-\sum_{\alpha\in B}s_{\alpha}$
for different integer fillings. Both $M_{\text{AF}}$ and $m_{\text{SLP}}$
vanish at $\nu=0,-2$ and closely follow each other at other integer
filling factors, as shown in Fig.~\ref{FIG:measurement}b, c. We see
weak SLP and AF order at $\nu=2$.

{\it Topological states}.---We next calculate the Chern number
of the flat bands through the eigenstates extracted from the self-consistent
density matrix, $\mathcal{C}_{n}(\sigma)=-(i/2\pi)\int_{\text{BZ}}\text{d}^{2}k\,\hat{n}\cdot[\nabla_{\mathbf{k}}\times\langle\phi_{\mathbf{k}}^{(n)}(\sigma)|\nabla_{\mathbf{k}}|\phi_{\mathbf{k}}^{(n)}(\sigma)\rangle],$
with the integral set over the 2D Brillouin zone oriented in the $\hat{n}$-axis.
In the panels of Fig.~\ref{FIG:band_structure} we depict the Chern number of the topological
flat bands with their corresponding spin polarization. At $\nu<0$ all bands in
red (spin $\sigma$) are topological, whereas the blue bands (spin $-\sigma$) are trivial.  For $\nu>0$ the blue bands become topological and 
the red ones trivial. All bands are trivial at $\nu=0$.
We find that the total Chern number of the occupied minibands with
$|\nu|=1,3$ is $\mathcal{C}=1$ for all values of $\varepsilon$
with insulating states. At integer fillings $|\nu|=0,2$ the occupied
mini-bands have zero net Chern number. 

We verify this result by doing
numerical measurements of the QAH order parameter through the real-space density matrix,
$
m_{\text{QAH}}=\frac{1}{3\sqrt{3}}\,\text{Im}\sum_{\langle\langle\alpha,\beta\rangle\rangle}\eta_{\alpha\beta}\hat{\rho}{}_{\alpha\beta}(0,\sigma),\label{eq:Maqh}
$
which probes for chiral loop currents among next-nearest-neighbor
sites $\langle\langle\alpha,\beta\rangle\rangle$, where $\eta_{\alpha\beta}=\pm1$
for clockwise or counterclockwise hopping in the honeycomb plaquette
of each monolayer \cite{Haldane}. As shown in Fig.~\ref{FIG:measurement}d, $m_{\text{QAH}}=0$
at even filling factors and is finite at odd fillings, in agreement
with the calculated Chern numbers depicted in Fig.~\ref{FIG:measurement}e.  We note that
the $\nu=-3$ state for $\varepsilon\geq18$ and $\nu=\pm1$ for $\varepsilon=30$ are  spin polarized metals
 \cite{Note2-1}.  Although we observe bulk loop currents in this state ($m_{\text{QAH}}\neq0$),
they do not generate chiral topological edge modes and the Chern
number is not well defined because of the finite Fermi surface around the $\Gamma$ point.

{\it Valley polarization}.--- In TBG the valleys 
are emergent quantum numbers that set the overall degeneracy of the
flat bands in the non-interacting limit \cite{Pereira,Dou}. To measure
VP effects we use the order parameter
$
m_{\text{VP}}=\frac{1}{3\sqrt{3}}\text{Im}\negmedspace\sum_{\langle\langle\alpha,\beta\rangle\rangle}\delta_{\alpha}\eta_{\alpha\beta}\hat{\rho}{}_{\alpha\beta}(0,\sigma),\label{eq:mV}
$
with $\delta_{\alpha}=\pm1$ for $\alpha\in A$ or $B$, which sums
over the difference between the loop currents in sublattices $A$
and $B$ in each monolayer \cite{Ramires}. In Fig.~\ref{FIG:measurement}f we show the
total VP calculated for integer fillings. The
behavior of $m_{\text{VP}}$ is not monotonic with the strength of
interactions. For all values of
$\varepsilon$ we find that VP and SLP are Pauli blocked at half filling,
$m_{\text{VP}}=m_{\text{SLP}}=0$, due to the $\nu=0$ state being
maximally spin polarized ($M_{\text{FM}}=4$), exhausting the occupation
of the flat bands with the same spin. 

{\it Discussion}.---Exact diagonalization results \cite{Potasz}
and exact calculations in the chiral limit of the continuum model
\cite{Bernevig} indicate that unrestricted HF may faithfully capture
the ground state of magic angle TBG at integer fillings, where correlation
effects are minimal, and interactions are mostly dominated by exchange.
Transport experiments indicate the prevalence
of insulating states at $\nu=\pm2,\,3$ \cite{Zondiner,Morissette}. Our results, summarized in Fig. 1, correctly reproduce the observed hierarchy of many-body gaps separating those states from the $\nu=0,\pm1,-3$ ones for $\varepsilon \gtrsim 24$. Our results also correctly show a strong particle-hole asymmetry between the $\nu=\pm3$ states, not captured in continuum models. 

We note that our results at $\nu=0$ strongly depart from prior quantum Monte Carlo and HF results in low energy effective models \cite{Liao, Hofmann,Liao, Gonzalez, Liu}, and indicate the presence of a pure spin ferromagnet with maximal polarization. The $\nu=\pm1,\pm3$ states are spin-valley polarized insulators with Chern number $\mathcal{C}=1$, whereas the $\nu=\pm2$ states are spin-valley polarized with zero net Chern number. The $\nu=\pm2$ results depart from prior exact diagonalization and HF calculations in low energy models,
which predicted either spin \cite{Gonzalez, Potasz} or valley polarized states \cite{M Xie}. 
On the other hand, our results at  $\nu=\pm3$ qualitatively agree with previous studies \cite{F Xie, Potasz, F Xie 2} and are consistent with zero-field transport
measurements in hexagonal-boron nitride non-aligned samples \cite{Stepanov-1}.  
We find that SLP is accompanied
with the emergence of AF order in the remote bands at $\nu=\pm 3,\pm1, 2$, and with sublattice
modulation of the ferromagnetically ordered state in the flat bands.
Recent electron spin resonance transport measurements have found evidence
of magnetism and AF order in magic angle TBG at $\nu=\pm2,3$ \cite{Morissette}.

Our numerical calculations also shed light into the MIT transition
of those insulating states. The $\nu=0$ insulator
has been observed in some transport experiments \cite{Cao1} but not in others
\cite{Zondiner,Morissette}. According to the scaling of the many-body
gaps shown in Fig.~\ref{FIG:phase_diagram}b, the typical transport gap  for
$\nu=-2$ is observed to be in the meV range in most HBN encapsulated samples,  consistently
with an effective dielectric constant $\varepsilon\sim25\pm5$, close to the $\nu=0$  state  MIT. We
predict that solid insulating behaviors at $\nu=0$ and $\nu=\pm1$
could be observed in suspended samples. Our results set the framework for the use of large-scale HF lattice calculations 
in moire heterostructures.

\begin{acknowledgments}
We acknowledge V. N. Kotov and E. Khalaf
for helpful discussions. K.\ A.\ and B.\ U.\  acknowledge NSF grant DMR-2024864
for support. We also acknowledge OSCER and MCSR supercomputing centers
for support. 
\end{acknowledgments}

\newpage


\section*{Supplementary material}

# Supplementary Material of "Strongly Interacting Phases in Twisted Bilayer Graphene at the Magic Angle"

Khagendra Adhikari,[1, *] Kangjun Seo,[2] K. S. D. Beach,[3] and Bruno Uchoa[1, †]

[1]*Department of Physics and Astronomy, University of Oklahoma, Norman, Oklahoma 73019, USA*
[2]*School of Electrical and Computer Engineering,*
*University of Oklahoma, Tulsa, Oklahoma 74135, USA*
[3]*Department of Physics and Astronomy, The University of Mississippi, University, Mississippi 38677, USA*

## TIGHT-BINDING CALCULATION

### Twisted bilayer graphene

We rotate the top layer by $+\theta/2$ and the bottom layer by $-\theta/2$ about the $z$-axis passing through $AA$ sites. For an arbitrary twist angle, commensuration between the layers requires that [1]

$$\cos(\theta^\circ) = \frac{3m^2 + 3ms + \frac{s^2}{2}}{3m^2 + 3ms + s^2},$$  (1)

with $m$ and $s$ integers. The choice $s = 1$ and $m = 0$ corresponds to a $60^o$ rotation ($AA$ stacked untwisted bilayer), whereas $s = 1$, $m = 30$ corresponds to $\theta \approx 1.085$, which is close to the magic angle. The number of sites per unit cell is given by

$$N_b = 4(3m^2 + 3ms + s^2),$$  (2)

with the factor of 4 corresponding to the two sub-lattices in each layer and the two layers. The period of the moiré unit cell is $L = a/[2\sin(\theta/2)]$, with $a = 2.46$Å the lattice constant corresponding to the carbon-carbon distance $a_0 = a/\sqrt{3}$.

In the rotated layers,

$$\boldsymbol{g}_i^{\text{bottom}} = R(-\theta/2)\boldsymbol{g}, \qquad \boldsymbol{g}_i^{\text{top}} = R(\theta/2)\boldsymbol{g}, \qquad R(\theta) = \begin{pmatrix} \cos\theta & \sin\theta \\ -\sin\theta & \cos\theta \end{pmatrix},$$  (3)

where $\boldsymbol{g}_1 = 2\pi(1, -1/\sqrt{3})/a$ and $\boldsymbol{g}_2 = 2\pi(0, 2/\sqrt{3})/a$ are the original reciprocal lattice vectors of the monolayer.

Reciprocal space moiré superlattice vectors can be calculated as

$$\boldsymbol{G}_i = \boldsymbol{g}_i^{\text{bottom}} - \boldsymbol{g}_i^{\text{top}}, \quad i \in \{1, 2\},$$  (4)

The corresponding real space moiré superlattice vectors are

$$\mathbf{T}_1 = m\boldsymbol{a}_1 + (m + s)\boldsymbol{a}_2, \qquad \mathbf{T}_2 = R(\pi/3)\mathbf{T}_1,$$  (5)

with $\boldsymbol{a}_1 = a(1, 0)$ and $\boldsymbol{a}_2 = a(1/2, \sqrt{3}/2)$ the generators of the original Bravais lattice. The atomic positions in the monolayer unit cell are $\boldsymbol{r}_A = (0, 0)$ and $\boldsymbol{r}_B = (a_1 + a_2)/3 = a(1/2, 1/2\sqrt{3})$, whereas the coordinates of the hexagonal center are $\boldsymbol{r}_{HC} = (2\boldsymbol{a}_1 - \boldsymbol{a}_2)/3$.

To calculate tight-binding band structure we follow ref. 3 and adopt the Slater-Koster parametrization

$$t(\delta r) = V_\sigma \cos^2\theta_z e^{-\frac{\delta r - d_0}{\lambda}} + V_\pi \sin^2\theta_z e^{-\frac{\delta r - a_0}{\lambda}},$$  (6)

where $\cos\theta_z = dz/\delta r$ with $\delta r = |\boldsymbol{r} - \boldsymbol{r}'|$ and $dz = |\boldsymbol{r}_z|$ [2]. $\lambda = 0.184a = 0.4525$Å is the $p$-orbital decay length, $V_\sigma = -0.48$eV is the interlayer hopping when two carbon atoms are exactly above each other, $V_\pi = 2.7$eV is the nearest-neighbors intralayer hopping strength, and $d_0 = 3.35$Å is the unrelaxed interlayer distance [3]. We implement the cutoff distance for hopping using a Heaviside step function $\Theta(5a_0 - r)$ including the first five nearest neighbor sites.

In the absence of corrugation the above parameters do not lead to a gap separating the flat bands from the remote ones (see Fig. 1). Corrugation due to lattice relaxation in the $z$ direction increases (reduces) the interlayer distance

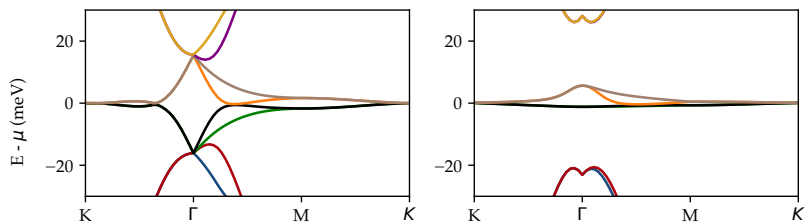

FIG. 1. Non-interacting tight-binding dispersion in the absence of lattice relaxation (left) and with relaxation (right), for optimized structure. Relaxation effects make the low energy bands flatter and stabilize a large gap of $\sim 20$meV with the remote bands.

in $AA(AB)$ site regions [4]. The lattice relaxation in which the $z-$coordinate of the carbon atoms is parametrized as

$$\boldsymbol{r}_z(\boldsymbol{r}_{xy}) = d_0 + d_1 \sum_{i=1,2,3} \cos\left(\boldsymbol{G}_i \cdot \boldsymbol{r}_{xy}\right) \tag{7}$$

where $\boldsymbol{G}_i \in \{\boldsymbol{G}_1, \boldsymbol{G}_2, \boldsymbol{G}_3 = (\boldsymbol{G}_1 + \boldsymbol{G}_2)\}$ are shortest three reciprocal lattice vectors, $d_0 = (d_{\mathrm{AA}}(\theta) + 2d_{\mathrm{AB}})/3$, $d_1 = 2(d_{\mathrm{AA}}(\theta) - d_{\mathrm{AB}})/9$. The twist-angle dependent displacement can be written as [4]

$$d_{\mathrm{AA}}(\theta) = d_{\mathrm{AB}} + (d_{\mathrm{AA}} - d_{\mathrm{AB}}) \exp\{-0.009\theta^2\} \tag{8}$$

with $d_{\mathrm{AA}} = 3.6$Å and $d_{\mathrm{AB}} = 3.35$Å are the relaxed inter-layer distance at $AA$ and $AB$ regions, respectively [3, 4].

## Coulomb Interaction

We consider Coulomb interactions screened by the metallic gates, consistently with the experimental setup. We use the following two-gated screened potential

$$U_1(\boldsymbol{r}) = U_\xi \sum_{m=-\infty}^{\infty} \frac{(-1)^m}{\sqrt{(r/\xi)^2 + m^2}}, \tag{9}$$

where $U_\xi = e^2/(\varepsilon\xi)$, $\xi \sim 10$ nm is the distance between the plates of the metallic contacts and $\varepsilon$ is the background dielectric constant. $U_\xi = 25$ meV for $\varepsilon = 6$ [5]. For $r > \xi$, the above equation can be simplified to [6, 7],

$$U_2(\boldsymbol{r}) = U_\xi \frac{2\sqrt{2}e^{-\pi r/\xi}}{\sqrt{r/\xi}}. \tag{10}$$

We use Eq. (9) for $r < \xi$ and Eq. (10) otherwise. In the opposite limit $r/\xi \ll 1$, we regularize the on site Coulomb interaction using a screened potential of the form [8]

$$U_3(\boldsymbol{r}) = \frac{e^2}{\varepsilon(0.116 + r)} eV = \frac{1.438 \, eV \, nm}{\varepsilon(0.116 + d)nm}. \tag{11}$$

We choose a smooth function that numerically interpolates between (9) and the on-site Hubbard term in (11) as follows

$$U(\boldsymbol{r}) = \begin{cases} \min(U_1(\boldsymbol{r}), U_3(\boldsymbol{r})), & \text{if } r < \xi \\ U_2(\boldsymbol{r}), & \text{otherwise} \, . \end{cases} \tag{12}$$

# HARTREE-FOCK CALCULATION

## Self-consistent loop

The full Hamiltonian is given by the sum of non-interacting and interacting terms as

$$
\begin{aligned}
\mathcal{H} &= \mathcal{H}_0 + \mathcal{H}_I \\
&= \sum_{\boldsymbol{k},\sigma} \left[ \mathcal{H}_0(\boldsymbol{k},\sigma) + \mathcal{H}^{\mathrm{F}}(\boldsymbol{k},\sigma) + \mathcal{H}^{\mathrm{H}}(\sigma) \right] \\
&= \sum_{\boldsymbol{k},\sigma} \mathcal{H}(\boldsymbol{k},\sigma),
\end{aligned}
\tag{13}
$$

where $\mathcal{H}^{\mathrm{H}}(\sigma)$ and $\mathcal{F}^{\mathrm{H}}(\boldsymbol{k},\sigma)$ are the Hartree and Fock contributions respectively. In unrestricted Hartree-Fock calculations, crystal momentum $\boldsymbol{k}$ and spin $\sigma$ are good quantum numbers. Therefore, the Hamiltonian matrix is block diagonal in the $\boldsymbol{k}$ and $\sigma$ spaces. The Hamiltonian matrix corresponding to a moire unit cell with $N_b$ sites has the form

$$
\mathcal{H}(\boldsymbol{k},\sigma) = 
\begin{array}{c}
\alpha_1 \\
\vdots \\
\alpha_{N_{\mathrm{b}}}
\end{array}
\begin{pmatrix}
\overset{\beta_1}{h_{11}(\boldsymbol{k},\sigma)} & \overset{\cdots}{\cdots} & \overset{\beta_{N_{\mathrm{b}}}}{h_{1N_{\mathrm{b}}}(\boldsymbol{k},\sigma)} \\
\vdots & \ddots & \vdots \\
h_{N_{\mathrm{b}}1}(\boldsymbol{k},\sigma) & \cdots & h_{N_{\mathrm{b}}N_{\mathrm{b}}}(\boldsymbol{k},\sigma)
\end{pmatrix}_{N_{\mathrm{b}} \times N_{\mathrm{b}}}.
\tag{14}
$$

We use a Monkhorst-Pack type $N_{\mathrm{u}} = N_1 \times N_2$ grid sampling of $\boldsymbol{k}$ momentum vectors in the rhombic Brillouin-zone [9]. The sampling of $\boldsymbol{k}$ vectors is restricted to certain discrete values as follows,

$$
\boldsymbol{k}_{m_1,m_2} = \frac{2m_1 - N_1 - 1}{2N_1}\boldsymbol{b}_1 + \frac{2m_2 - N_2 - 1}{2N_2}\boldsymbol{b}_2,
\tag{15}
$$

where $m_i \in \{1, 2, \cdots, N_i\}$ ($i = 1, 2$). $\boldsymbol{b}_1$ and $\boldsymbol{b}_2$ are moiré lattice vectors in the reciprocal space. We use a grid size $N_1 = N_2 = 9$ with an odd number of $k$ points, such that the $\Gamma$ and $K$ points are included in the $k$-mesh. In this prescription, the $k$-mesh preserves the hexagonal symmetry of the lattice.

We diagonalize each block matrix $\mathcal{H}(\boldsymbol{k},\sigma)$ individually and get $N_{\mathrm{b}}$ eigenvalues and corresponding eigenvectors for each $\boldsymbol{k}$. Diagonalization of $\mathcal{H}(\boldsymbol{k},\sigma)$ gives $N_{\mathrm{b}}$ eigenvalues $E_n(\boldsymbol{k},\sigma)$. The $n$-th column of the matrix

$$
\Phi(\boldsymbol{k},\sigma) = 
\begin{array}{c}
\alpha_1 \\
\vdots \\
\alpha_{N_{\mathrm{b}}}
\end{array}
\begin{pmatrix}
\overset{n_1}{\phi_1^1(\boldsymbol{k},\sigma)} & \overset{\cdots}{\cdots} & \overset{n_{N_{\mathrm{b}}}}{\phi_1^{N_{\mathrm{b}}}(\boldsymbol{k},\sigma)} \\
\vdots & \ddots & \vdots \\
\phi_{N_{\mathrm{b}}}^1(\boldsymbol{k},\sigma) & \cdots & \phi_{N_{\mathrm{b}}}^{N_{\mathrm{b}}}(\boldsymbol{k},\sigma)
\end{pmatrix}_{N_{\mathrm{b}} \times N_{\mathrm{b}}}
\tag{16}
$$

gives the corresponding eigenvector of the $n$-th band, with $n \in \{1, 2, \cdots, N_{\mathrm{b}}\}$. The normalized eigenvectors $\phi_\alpha^n(\boldsymbol{k},\sigma)$ form a complete basis and satisfy

$$
\sum_\alpha \phi_\alpha^{n*}(\boldsymbol{k},\sigma)\phi_\alpha^{n'}(\boldsymbol{k},\sigma') = \delta_{nn'}\delta_{\sigma\sigma'},
\tag{17}
$$

and

$$
\sum_n \phi_\alpha^{n*}(\boldsymbol{k},\sigma)\phi_\beta^n(\boldsymbol{k},\sigma') = \delta_{\alpha\beta}\delta_{\sigma\sigma'}.
\tag{18}
$$

The number of crystal momenta $N_{\mathrm{u}}$ sampled within the first Brillouin zone is equivalent to the number of unit cells in the real space lattice. The system can accommodate a maximum of $N_{\mathrm{s}}$ spin-$\sigma$ electrons. In total, there are $N_{\mathrm{s}} = N_{\mathrm{u}} \times N_{\mathrm{b}}$ eigenvalues per spin in a system containing $N_{\mathrm{s}}$ sites in total. We order $2N_{\mathrm{s}}$ states (combining both spin types for all $\boldsymbol{k}$ vectors) in the ascending order of eigenvalues.

The chemical potential $\mu$ is the eigenvalue corresponding to the topmost filled state for a given total number of particles in the system, $N = (N_{\mathrm{b}} + \nu) \times N_{\mathrm{u}} \leq 2N_{\mathrm{s}}$, with $N_{\mathrm{b}}$ the number of sites/atoms per unit cell. In other words, $\mu$ is solved self-consistently at a constant electron density $\nu$, which is fixed at each self-consistent iteration by solving

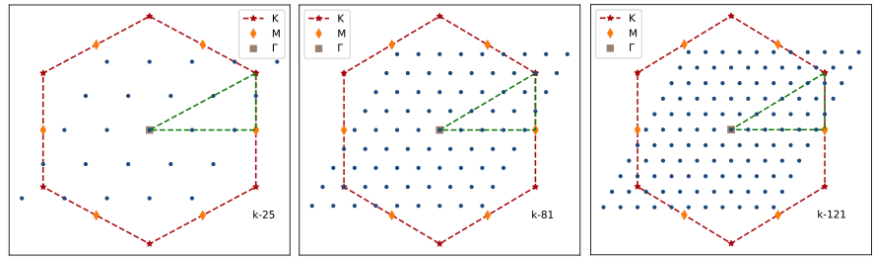

FIG. 2. Coarse graining of momentum in the BZ through a $k$-mesh with (left to right) 25, 81 and 121 points. Green dashed line: irreducible BZ. All cases include the $\Gamma$ point. The $9 \times 9$ mesh (81 points) we used in the calculation also includes the $K$ points.

the equation

$$\frac{1}{N_u} \sum_{\boldsymbol{k}, \sigma, n} \theta(\mu - \epsilon_{\boldsymbol{k}, \sigma, n}) = N_{\mathrm{a}} + \nu. \tag{19}$$

Once we get chemical potential $\mu$, the density matrix in Eq. (29) is calculated by summing over all the occupied states. Hence, the density matrix $\rho(\boldsymbol{R}, \sigma)$ is the only mean-field parameter in the self-consistent Hartree-Fock calculation. We implement the strong convergence criteria for the density matrix where the root mean square of the difference in density matrices with the previous iteration is less than the tolerance, *i.e.* $(\rho(0, \sigma)_{\mathrm{curr}} - \rho(0, \sigma)_{\mathrm{prev}}) < 10^{-8}$. We use linear mixing for the density matrix,

$$\rho(\boldsymbol{R}, \sigma)_{\mathrm{prev}} = X \, \rho(\boldsymbol{R}, \sigma)_{\mathrm{curr}} + (1 - X) \, \rho(\boldsymbol{R}, \sigma)_{\mathrm{prev}}, \tag{20}$$

with mixing coefficient $X = 0.75$. We use this density matrix to build the Hartree and Fock matrices for the next iteration.

### Convergence test

The momenta in the BZ is coarse grained over a $k$-mesh. In order to verify convergence and stability of the calculated self-consistent density matrix with the number of points of the $k$-mesh, we tested the method considering three different meshes with 25, 81 and 121 points. All tested meshes included the $\Gamma$ point, whereas the $9 \times 9$ mesh (81 points) also included the $K$ points, as shown in Fig. 2. In all meshes we tested, we did not observe any differences in the many-body gaps or in the measurements for SP and QAH states.

The measurements of VP, SLP and AF involve differences of density matrix measurements between different sublattices and are more sensitive to the coarse graining and numerical noise. We observed small differences in the measurements of SLP, VP and AF with different meshes and seeds of random noise in the initial density matrix. The $9 \times 9$ mesh was found to be numerically stable in all measurements for different seeds we tested. We did not observe significant numerical improvements in the denser mesh with 121 points compared to the one with 81 points, which was used in all numerical calculations shown in the main text.

### Interaction matrix

We consider a four-fermion interaction term for the Coulomb interaction, has no explicit spin dependence:

$$\mathcal{H}_I = \frac{1}{2} \sum_{\alpha\beta} \sum_{ij} \sum_{\sigma\sigma'} U_{\alpha\beta}(\boldsymbol{R}_j - \boldsymbol{R}_i) d_{\alpha, i, \sigma}^{\dagger} d_{\beta, j, \sigma'}^{\dagger} d_{\beta, j, \sigma'} d_{\alpha, i, \sigma}, \tag{21}$$

where

$$U_{\alpha\beta}(\boldsymbol{R}_j - \boldsymbol{R}_i) \equiv U(\boldsymbol{R}_j + \boldsymbol{\tau}_\beta - \boldsymbol{R}_i - \boldsymbol{\tau}_\alpha), \tag{22}$$

with $\tau_\alpha$ the position of lattice site $\alpha$ with respect to the center of the unit cell located at position $\boldsymbol{R}_i$. The Fourier transform of (21) is

$$V = \frac{1}{2} \sum_{\sigma\sigma'} \sum_{\boldsymbol{k},\boldsymbol{k}'} \sum_{\boldsymbol{q}} \sum_{\alpha\beta} U_{\alpha\beta}(\boldsymbol{q}) d^\dagger_{\boldsymbol{k},\alpha,\sigma} d^\dagger_{\boldsymbol{k}',\beta,\sigma'} d_{\boldsymbol{k}'-\boldsymbol{q},\beta,\sigma'} d_{\boldsymbol{k}+\boldsymbol{q},\alpha,\sigma},$$ (23)

with

$$\begin{aligned} U_{\alpha\beta}(\boldsymbol{q}) &= \frac{1}{N_u} \sum_{\boldsymbol{R}} U(|\boldsymbol{R} + \tau_\beta - \tau_\alpha|) e^{i\boldsymbol{q}\cdot\boldsymbol{R}} \\ &= \frac{1}{N_u} \sum_{\boldsymbol{R}} U_{\alpha\beta}(\boldsymbol{R}) e^{i\boldsymbol{q}\cdot\boldsymbol{R}} \end{aligned}$$ (24)

the Fourier transform of the Coulomb interaction. The Bravais superlattice vectors $\boldsymbol{R} = \boldsymbol{R}_j - \boldsymbol{R}_i$ are calculated using real space lattice vectors $\boldsymbol{a}_1$ and $\boldsymbol{a}_1$ as follows

$$R_{m_1,m_2} = m_1\boldsymbol{a}_1 + m_2\boldsymbol{a}_2, \qquad m_i \in \{-L/2, \cdots, 0, \cdots, L/2\}$$ (25)

and the Coulomb interaction is calculated using a list $\boldsymbol{R}_i = \{R_{m_1,m_2}\}$ of Bravais lattice sites.

The interaction satisfies $U_{\alpha\beta}(\boldsymbol{R}) = U_{\beta\alpha}(-\boldsymbol{R})$, and $\hat{U}(\boldsymbol{R}) = \hat{U}^T(-\boldsymbol{R})$, and hence $U_{\alpha\beta}(\boldsymbol{R})$ is not a Hermitian matrix. On the other hand, its Fourier transform $U_{\alpha\beta}(\boldsymbol{q})$ is Hermitian,

$$U^\dagger_{\alpha\beta}(\boldsymbol{q}) = U_{\beta\alpha}(-\boldsymbol{q}) = \frac{1}{N_u} \sum_{\boldsymbol{R}} U_{\beta\alpha}(\boldsymbol{R}) e^{-i\boldsymbol{q}\cdot\boldsymbol{R}} = \frac{1}{N_u} \sum_{\boldsymbol{R}} U_{\beta\alpha}(-\boldsymbol{R}) e^{i\boldsymbol{q}\cdot\boldsymbol{R}} = \frac{1}{N_u} \sum_{\boldsymbol{R}} U_{\alpha\beta}(\boldsymbol{R}) e^{i\boldsymbol{q}\cdot\boldsymbol{R}} = U_{\alpha\beta}(\boldsymbol{q}).$$ (26)

Besides, $\bar{U}^*_{\alpha\beta}(\boldsymbol{q}) = \bar{U}_{\alpha\beta}(-\boldsymbol{q})$, since $U_{\alpha\beta}(\boldsymbol{R})$ is real.

### Mean-field decomposition

The expectation value of $d$-fermions in the Hartree-Fock mean-field basis can be written as

$$\begin{aligned} \langle d^\dagger_{\boldsymbol{k}\sigma,\alpha} d_{\boldsymbol{k}'\sigma',\beta} \rangle &\equiv \langle \mathrm{MF} | d^\dagger_{\boldsymbol{k}\sigma,\alpha} d_{\boldsymbol{k}'\sigma',\beta} | \mathrm{MF} \rangle \\ &= \sum_n \phi^{n*}_\alpha(\boldsymbol{k},\sigma) \phi^n_\beta(\boldsymbol{k}',\sigma') \theta(\mu - \epsilon_{\boldsymbol{k},n}) \delta_{\boldsymbol{k}\boldsymbol{k}'} \delta_{\sigma\sigma'} \\ &= \sum_n^{\mathrm{occ}} \phi^{n*}_\alpha(\boldsymbol{k},\sigma) \phi^n_\beta(\boldsymbol{k}',\sigma') \delta_{\boldsymbol{k}\boldsymbol{k}'} \delta_{\sigma\sigma'}, \end{aligned}$$ (27)

summed over all the occupied bands. From Eq. (27), the momentum space density matrix can be written as

$$\begin{aligned} \bar{\rho}_{\alpha\beta}(\boldsymbol{k},\sigma) &= \langle \mathrm{MF} | d^\dagger_{\boldsymbol{k}\sigma,\beta} d_{\boldsymbol{k}\sigma,\alpha}^n | \mathrm{MF} \rangle \\ &= \sum_n^{\mathrm{occ}} \phi^{n*}_\beta(\boldsymbol{k},\sigma) \phi^n_\alpha(\boldsymbol{k},\sigma). \end{aligned}$$ (28)

Here, $\bar{\rho}(\boldsymbol{k},\sigma) = \bar{\rho}^*(-\boldsymbol{k},\sigma)$. The real-space density matrix is defined as

$$\begin{aligned} \rho_{\alpha\beta}(\boldsymbol{R},\sigma) &= \frac{1}{N_u} \sum_{\boldsymbol{k}} \sum_n^{\mathrm{occ}} e^{-i\boldsymbol{k}\cdot\boldsymbol{R}} \phi^{n*}_\beta(\boldsymbol{k},\sigma) \phi^n_\alpha(\boldsymbol{k},\sigma) \\ &= \frac{1}{N_u} \sum_{\boldsymbol{k}} e^{-i\boldsymbol{k}\cdot\boldsymbol{R}} \bar{\rho}_{\alpha\beta}(\boldsymbol{k},\sigma) \end{aligned}$$ (29)

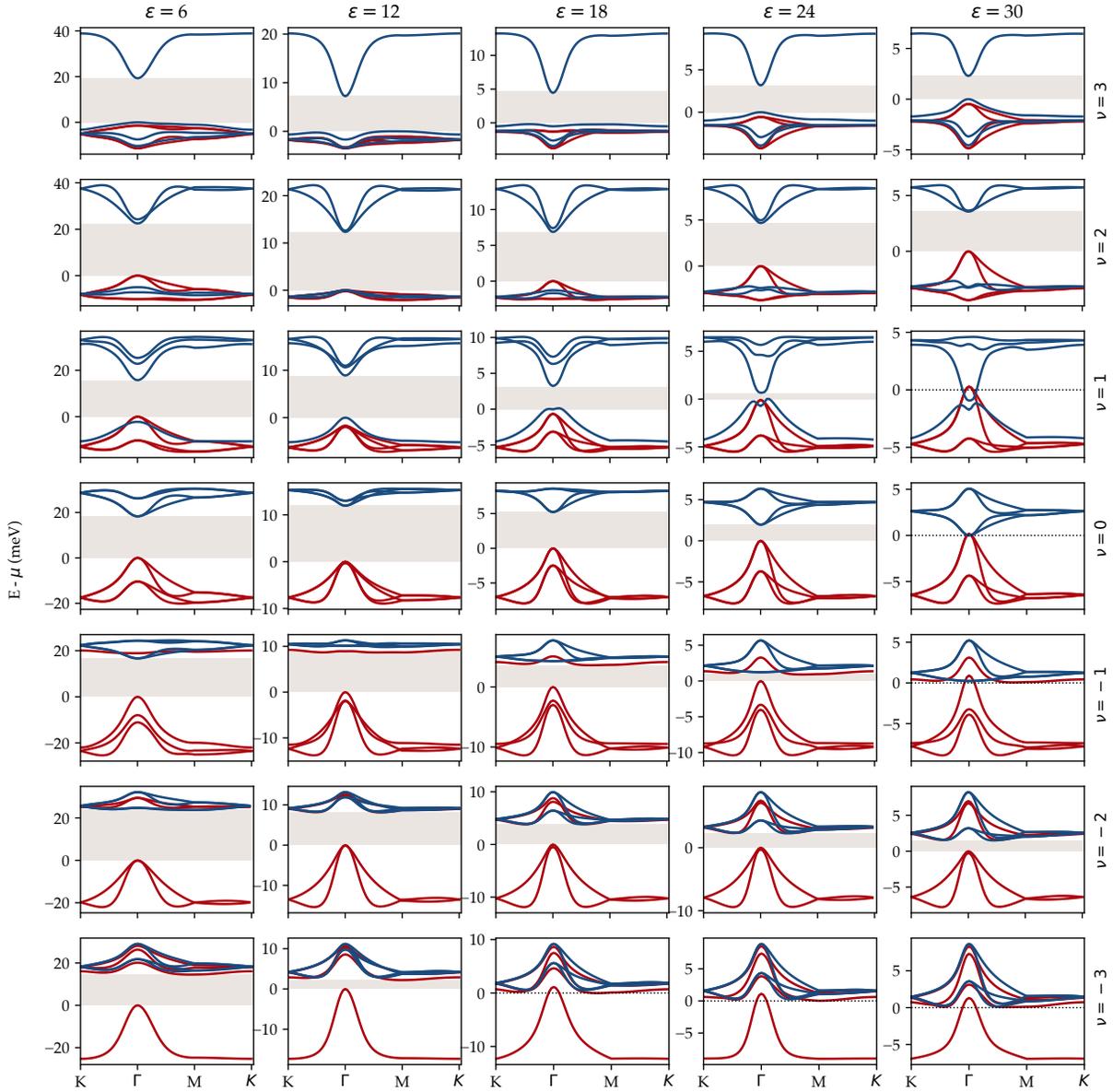

FIG. 3. Band structure of the low energy bands for dielectric constants $\varepsilon = 6, 12, 18, 24$ and $30$ at integer filling factors, from $\nu = 3$ to $\nu = -3$ (top to bottom). Red lines: spin $\sigma$; blue lines: spin $-\sigma$. The gray region indicates the many-body transport gap. All bands in red (blue) are topological for $\nu < 0$ ($\nu > 0$) and trivial otherwise. The Chern numbers of all occupied bands add up to 1 at odd fillings and to 0 at even fillings. Energies measured away from the chemical potential $\mu$.

where the diagonal element $\rho_\alpha(0, \sigma)$ is the average particle density of spin $\sigma$ at site $\alpha$. The trace $\mathrm{Tr}\,(\rho_{\alpha\beta}(0, \sigma))$ gives the average number of spin $\sigma$ particles per unit cell. From the general properties of the density matrix [10],

$$\bar{\rho}_{\alpha\beta}(\boldsymbol{k}, \sigma) = \bar{\rho}^*_{\beta\alpha}(\boldsymbol{k}, \sigma) = \bar{\rho}_{\beta\alpha}(-\boldsymbol{k}, \sigma) = \bar{\rho}^*_{\alpha\beta}(-\boldsymbol{k}, \sigma), \tag{30}$$

and

$$\rho_{\alpha\beta}(\boldsymbol{R}, \sigma) = \rho^\dagger_{\beta\alpha}(-\boldsymbol{R}, \sigma), \tag{31}$$

then $\rho(\boldsymbol{R}, \sigma) = \rho^\dagger(-\boldsymbol{R}, \sigma)$. I.e, $\rho(0, \sigma)$ is an Hermitian matrix. The density matrix $\rho_{\alpha\beta}(\boldsymbol{R}, \sigma)$ is used as a self-consistency parameter in the numerical calculation. It is diagonal if all the states are fully occupied with $N = 2 * N_\mathrm{s}$ (or $\nu = N_\mathrm{b}$).

In the mean-field decomposition, Coulomb interaction in Eq. (23) can be separated into direct (Hartree) and exchange (Fock) terms,

$$
\begin{aligned}
\mathcal{H}_I^{\mathrm{H}} &\approx \frac{1}{2} \sum_{\sigma\sigma'} \sum_{\boldsymbol{k},\boldsymbol{k}',\boldsymbol{q}} \sum_{\alpha\beta} \left[ \bar{U}_{\alpha\beta}(\boldsymbol{q}) d_{\boldsymbol{k}\sigma,\alpha}^{\dagger} d_{\boldsymbol{k}+\boldsymbol{q}\sigma,\alpha} \langle d_{\boldsymbol{k}'\sigma',\beta}^{\dagger} d_{\boldsymbol{k}'+\boldsymbol{q}\sigma',\beta} \rangle + \bar{U}_{\alpha\beta}(\boldsymbol{q}) \langle d_{\boldsymbol{k}\sigma,\alpha}^{\dagger} d_{\boldsymbol{k}+\boldsymbol{q}\sigma,\alpha} \rangle d_{\boldsymbol{k}'\sigma',\beta}^{\dagger} d_{\boldsymbol{k}'-\boldsymbol{q}\sigma',\beta} \right] \\
&= \sum_{\sigma,\sigma'} \sum_{\boldsymbol{k},\boldsymbol{k}'} \sum_{\alpha\beta} \sum_{n}^{\mathrm{occ}} \bar{U}_{\alpha\beta}(0)\, \phi_{\beta}^{n*}(\boldsymbol{k}',\sigma') \phi_{\beta}^{n}(\boldsymbol{k}',\sigma') d_{\boldsymbol{k}\sigma,\alpha}^{\dagger} d_{\boldsymbol{k}\sigma,\alpha} \\
&= \sum_{\sigma,\sigma'} \sum_{\boldsymbol{k},\boldsymbol{k}'} \sum_{\alpha\beta} \sum_{n}^{\mathrm{occ}} \frac{1}{N_{\mathrm{u}}} \sum_{\boldsymbol{R}} U_{\alpha\beta}(\boldsymbol{R})\, \phi_{\beta}^{n*}(\boldsymbol{k}',\sigma') \phi_{\beta}^{n}(\boldsymbol{k}',\sigma') d_{\boldsymbol{k}\sigma,\alpha}^{\dagger} d_{\boldsymbol{k}\sigma,\alpha} \\
&= \sum_{\sigma,\boldsymbol{k},\alpha} h_{\alpha}^{\mathrm{H}}(\sigma) d_{\boldsymbol{k}\sigma,\alpha}^{\dagger} d_{\boldsymbol{k}\sigma,\alpha}
\end{aligned}
\tag{32}
$$

and

$$
\begin{aligned}
\mathcal{H}_I^{\mathrm{F}} &\approx -\frac{1}{2} \sum_{\sigma\sigma'} \sum_{\boldsymbol{k},\boldsymbol{k}',\boldsymbol{q}} \sum_{\alpha\beta} \left[ \bar{U}_{\alpha\beta}(\boldsymbol{q}) d_{\boldsymbol{k}\sigma,\alpha}^{\dagger} d_{\boldsymbol{k}'-\boldsymbol{q}\sigma',\beta} \langle d_{\boldsymbol{k}'\sigma',\beta}^{\dagger} d_{\boldsymbol{k}+\boldsymbol{q}\sigma,\alpha} \rangle + \bar{U}_{\alpha\beta}(\boldsymbol{q}) \langle d_{\boldsymbol{k}\sigma,\alpha}^{\dagger} d_{\boldsymbol{k}'-\boldsymbol{q}\sigma',\beta} \rangle d_{\boldsymbol{k}'\sigma',\beta}^{\dagger} d_{\boldsymbol{k}+\boldsymbol{q}\sigma,\alpha} \right] \\
&= -\sum_{\sigma} \sum_{\boldsymbol{k}} \sum_{\alpha\beta} \sum_{\boldsymbol{k}'} \sum_{n}^{\mathrm{occ}} \bar{U}_{\alpha\beta}(\boldsymbol{k}'-\boldsymbol{k})\, \phi_{\beta}^{n*}(\boldsymbol{k}',\sigma) \phi_{\alpha}^{n}(\boldsymbol{k}',\sigma) d_{\boldsymbol{k}\sigma,\alpha}^{\dagger} d_{\boldsymbol{k}\sigma,\beta} \\
&= -\sum_{\sigma} \sum_{\boldsymbol{k}} \sum_{\alpha\beta} \sum_{\boldsymbol{k}'} \sum_{n}^{\mathrm{occ}} \frac{1}{N_{\mathrm{u}}} \sum_{\boldsymbol{R}} e^{-i(\boldsymbol{k}'-\boldsymbol{k})\cdot\boldsymbol{R}} U_{\alpha\beta}(\boldsymbol{R})\, \phi_{\beta}^{n*}(\boldsymbol{k}',\sigma) \phi_{\alpha}^{n}(\boldsymbol{k}',\sigma) d_{\boldsymbol{k}\sigma,\alpha}^{\dagger} d_{\boldsymbol{k}\sigma,\beta} \\
&= -\sum_{\sigma,\boldsymbol{k}} \sum_{\alpha\beta} h_{\alpha\beta}^{\mathrm{F}}(\boldsymbol{k},\sigma) d_{\boldsymbol{k}\sigma,\alpha}^{\dagger} d_{\boldsymbol{k}\sigma,\beta}
\end{aligned}
\tag{33}
$$

where

$$
h_{\alpha}^{\mathrm{H}}(\sigma) = \sum_{\beta,\sigma'} \sum_{\boldsymbol{R}} U_{\alpha\beta}(\boldsymbol{R}) \rho_{\beta}(0,\sigma'),
\tag{34}
$$

and

$$
h_{\alpha\beta}^{\mathrm{F}}(\boldsymbol{k},\sigma) = \sum_{\boldsymbol{R}} e^{i\boldsymbol{k}\cdot\boldsymbol{R}} U_{\alpha\beta}(\boldsymbol{R}) \rho_{\alpha\beta}(\boldsymbol{R},\sigma)
\tag{35}
$$

are the Hartree and Fock matrix elements, respectively. All densities are measured away from the neutrality point. We subtract a uniform background of charge from the density matrix, $\rho_{\alpha,\beta}(\boldsymbol{R},\sigma) \to \rho_{\alpha,\beta}(\boldsymbol{R},\sigma) - \frac{1}{2} \delta_{\alpha,\beta} \delta_{\boldsymbol{R},0}$.

Once the real space density matrix is found to converge, the band structure of the many-body problem is extracted from the diagonalization of the Hartree-Fock Hamiltonian and calculated over a $k$-path passing through high symmetry points. The band structure of the low energy bands for $\varepsilon = 6, 12, 18, 24$ and $30$ are shown in Fig. 3 of the SM for integer filling factors $\nu \in [-3, 3]$.

### Numerical measurements

We performed numerical measurements using the density matrix, as described in the main text. In table I we report the contribution from both the low energy flat bands (FB) and remote bands (RB) to the different order parameters for $\varepsilon = 6, 12, 18, 24$ and $30$. We find that the spin polarization (SP) describes a ferromagnetic state that is restricted to the flat bands, with the remote bands having zero net polarization. Antiferromagnetism (AF) and sublattice polarization (SLP) measurements in the low energy flat bands are correlated (see table I), indicating the presence of spin modulation in the flat bands with the period of a lattice constant. With the exception of $\nu = 0$ and $-2$, we observe that all other integer fillings are sublattice polarized. SLP also correlates with the emergence of a staggered AF order in the remote bands.

The pattern of charge distribution and spin polarization of the flat bands in the moire unit cell are shown in Fig.

Order parameters for $\varepsilon = 6$

| ν | −3 | | −2 | | −1 | | 0 | | 1 | | 2 | | 3 | |
|---|---|---|---|---|---|---|---|---|---|---|---|---|---|---|
| OP | RB | FB | RB | FB | RB | FB | RB | FB | RB | FB | RB | FB | RB | FB |
| SP | 0.00 | 1.00 | 0.00 | 2.00 | 0.00 | 3.00 | 0.00 | 4.00 | 0.00 | 3.00 | 0.00 | 2.00 | 0.00 | 1.00 |
| AF | 0.16 | 0.12 | 0.00 | 0.00 | 0.20 | 0.15 | 0.00 | 0.00 | 0.26 | 0.20 | 0.06 | 0.04 | 0.20 | 0.16 |
| SLP | 0.08 | 0.12 | 0.00 | 0.00 | 0.10 | 0.15 | 0.00 | 0.00 | 0.12 | 0.20 | 0.03 | 0.04 | 0.10 | 0.16 |
| QAH | 0.17 | 0.22 | 0.00 | 0.00 | 0.17 | 0.23 | 0.00 | 0.00 | 0.16 | 0.23 | 0.00 | 0.00 | 0.16 | 0.23 |
| VP | 0.00 | 0.13 | 0.00 | 0.01 | 0.00 | 0.15 | 0.00 | 0.00 | 0.00 | 0.20 | 0.00 | 0.22 | 0.00 | 0.16 |

Order parameters for $\varepsilon = 12$

| ν | −3 | | −2 | | −1 | | 0 | | 1 | | 2 | | 3 | |
|---|---|---|---|---|---|---|---|---|---|---|---|---|---|---|
| OP | RB | FB | RB | FB | RB | FB | RB | FB | RB | FB | RB | FB | RB | FB |
| SP | 0.00 | 1.00 | 0.00 | 2.00 | 0.00 | 3.00 | 0.00 | 4.00 | 0.00 | 3.00 | 0.00 | 2.00 | 0.00 | 1.00 |
| AF | 0.07 | 0.12 | 0.00 | 0.00 | 0.11 | 0.20 | 0.00 | 0.00 | 0.09 | 0.16 | 0.01 | 0.02 | 0.07 | 0.14 |
| SLP | 0.04 | 0.12 | 0.00 | 0.00 | 0.07 | 0.20 | 0.00 | 0.00 | 0.05 | 0.16 | 0.01 | 0.02 | 0.04 | 0.14 |
| QAH | 0.07 | 0.18 | 0.00 | 0.00 | 0.07 | 0.19 | 0.00 | 0.00 | 0.07 | 0.19 | 0.00 | 0.00 | 0.07 | 0.19 |
| VP | 0.00 | 0.14 | 0.00 | 0.28 | 0.00 | 0.23 | 0.00 | 0.00 | 0.00 | 0.18 | 0.00 | 0.32 | 0.00 | 0.16 |

Order parameters for $\varepsilon = 18$

| ν | −3 | | −2 | | −1 | | 0 | | 1 | | 2 | | 3 | |
|---|---|---|---|---|---|---|---|---|---|---|---|---|---|---|
| OP | RB | FB | RB | FB | RB | FB | RB | FB | RB | FB | RB | FB | RB | FB |
| SP | 0.00 | 1.00 | 0.00 | 2.00 | 0.00 | 3.00 | 0.00 | 4.00 | 0.00 | 3.00 | 0.00 | 2.00 | 0.00 | 1.00 |
| AF | 0.03 | 0.08 | 0.00 | 0.00 | 0.05 | 0.14 | 0.00 | 0.00 | 0.03 | 0.09 | 0.01 | 0.02 | 0.05 | 0.13 |
| SLP | 0.02 | 0.08 | 0.00 | 0.00 | 0.03 | 0.14 | 0.00 | 0.00 | 0.02 | 0.09 | 0.01 | 0.02 | 0.03 | 0.13 |
| QAH | 0.03 | 0.12 | 0.00 | 0.00 | 0.04 | 0.17 | 0.00 | 0.00 | 0.03 | 0.11 | 0.00 | 0.00 | 0.04 | 0.17 |
| VP | 0.00 | 0.15 | 0.00 | 0.20 | 0.00 | 0.18 | 0.00 | 0.00 | 0.00 | 0.18 | 0.00 | 0.16 | 0.00 | 0.18 |

Order parameters for $\varepsilon = 24$

| ν | −3 | | −2 | | −1 | | 0 | | 1 | | 2 | | 3 | |
|---|---|---|---|---|---|---|---|---|---|---|---|---|---|---|
| OP | RB | FB | RB | FB | RB | FB | RB | FB | RB | FB | RB | FB | RB | FB |
| SP | 0.00 | 1.00 | 0.00 | 2.00 | 0.00 | 3.00 | 0.00 | 4.00 | 0.00 | 3.00 | 0.00 | 2.00 | 0.00 | 1.00 |
| AF | 0.03 | 0.12 | 0.00 | 0.00 | 0.02 | 0.09 | 0.00 | 0.00 | 0.01 | 0.04 | 0.01 | 0.02 | 0.03 | 0.12 |
| SLP | 0.02 | 0.12 | 0.00 | 0.00 | 0.01 | 0.09 | 0.00 | 0.00 | 0.01 | 0.04 | 0.01 | 0.02 | 0.02 | 0.12 |
| QAH | 0.03 | 0.15 | 0.00 | 0.00 | 0.03 | 0.16 | 0.00 | 0.00 | 0.01 | 0.05 | 0.00 | 0.00 | 0.03 | 0.15 |
| VP | 0.00 | 0.18 | 0.00 | 0.09 | 0.00 | 0.12 | 0.00 | 0.00 | 0.00 | 0.19 | 0.00 | 0.08 | 0.00 | 0.19 |

Order parameters for $\varepsilon = 30$

| ν | −3 | | −2 | | −1 | | 0 | | 1 | | 2 | | 3 | |
|---|---|---|---|---|---|---|---|---|---|---|---|---|---|---|
| OP | RB | FB | RB | FB | RB | FB | RB | FB | RB | FB | RB | FB | RB | FB |
| SP | 0.00 | 1.00 | 0.00 | 2.00 | 0.00 | 3.00 | 0.00 | 3.95 | 0.00 | 2.95 | 0.00 | 2.00 | 0.00 | 1.00 |
| AF | 0.02 | 0.11 | 0.00 | 0.00 | 0.02 | 0.07 | 0.00 | 0.00 | 0.00 | 0.02 | 0.01 | 0.02 | 0.02 | 0.09 |
| SLP | 0.02 | 0.11 | 0.00 | 0.00 | 0.01 | 0.07 | 0.00 | 0.00 | 0.00 | 0.02 | 0.01 | 0.02 | 0.01 | 0.09 |
| QAH | 0.02 | 0.13 | 0.00 | 0.00 | 0.02 | 0.14 | 0.00 | 0.00 | 0.00 | 0.03 | 0.00 | 0.00 | 0.02 | 0.14 |
| VP | 0.00 | 0.18 | 0.00 | 0.04 | 0.00 | 0.11 | 0.00 | 0.00 | 0.00 | 0.14 | 0.00 | 0.02 | 0.00 | 0.17 |

TABLE I. Contributions from the flat bands (FB) and remote bands (RB) to the order parameters. SP: spin polarization; AF: antiferromagnetism. SLP: sublattice polarization; QAH: quantum anomalous Hall; VP: valley polarization.

4. The charge distribution shown in panel 4a accumulates at the corners of the unit cell, near the $AA$ site regions. This region (orange diamond) is shown in detail in panel 4b. The spin polarization pattern is shown in Fig. 4c. The red diamond on the corner is shown in detail in panel 4d. The spin polarization has a sublattice modulation that becomes more pronounced near $AA$ sites. This pattern of polarization is absent at half and quarter filling ($\nu = 0$ and $\nu = -2$), but was observed in the insulating phases at all other integer fillings.

Our measurements of loop currents on the lattice indicates the emergence of a quantum anomalous Hall (QAH) state at odd filling factors. Interestingly, we find contributions to QAH order parameter not only in the flat bands, but also in some of the remote bands, which can be topological. At even filling factors we measured zero net contribution to the QAH order parameter both in the flat bands as in the remote ones. The detailed results of the measurements can be found on table I.

We note that at $\nu = -3$ the system undergoes a MIT below $\varepsilon = 18$, where the many-body transport gap closes, even though a finite optical gap is still present at all momenta in the BZ. The band structure shown in Fig. 3 (bottom panel on the right) indicates that this state describes a spin polarized metal with a finite Fermi surface around the $\Gamma$

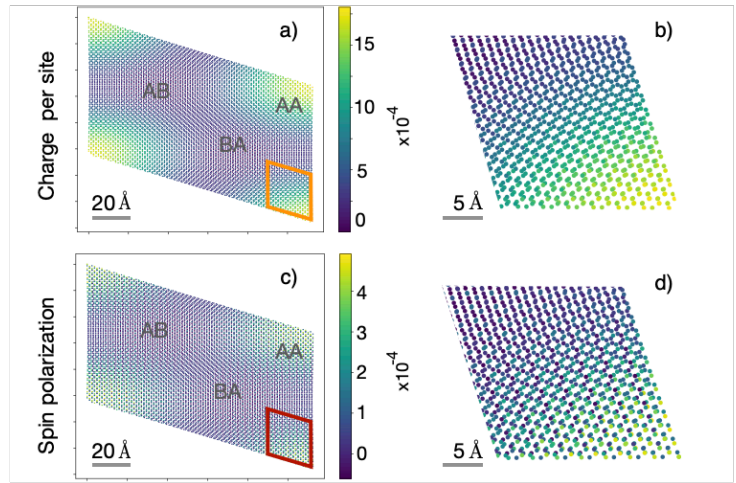

FIG. 4. Charge and spin distribution on the Moiré unit cell in magic angle TBG for $\varepsilon = 6$ and $\nu = 3$. a) Charge per site $\hat{\rho}_{\alpha\alpha}(0, \sigma)$ of the low energy flat bands, concentrated near $AA$ regions, at the corners on the unit cell (orange diamond), shown in b). The zoomed charge pattern shows sublattice symmetry breaking (SLP order). c) Spin polarization favors ferromagnetism, with amplitude modulation at the sublattice scale near $AA$ sites (red diamond), as shown in d).

point. The measurements of the QAH order parameter are non-zero, revealing loop currents in the bulk. While the corresponding edge states are in principle allowed to form inside the optical gap, they do not translate in topologically protected chiral edge states immune to backscattering effects, since the system has a Fermi surface. In that sense, a Chern number is not well defined. The same applies to the metallic states at $\nu = \pm 1$ for $\varepsilon = 30$, which have a finite QAH order parameter, indicating the presence of chiral loop currents in bulk, but no well defined Chern number.

We verified the topological nature of the flat bands by explicitly calculating their Chern number. At negative fillings ($\nu < 0$) the spin $\sigma$ bands (red lines in Fig. 3) are topological whereas the spin $-\sigma$ ones (blue) are topologically trivial. This pattern reverses at positive fillings ($\nu > 0$), with the spin $\sigma$ bands becoming trivial and the spin $-\sigma$ bands (blue) topological. We numerically observe that inversion symmetry is broken between the eigenvalues $E(\mathbf{k})$ and $E(-\mathbf{k})$ of all topological flat bands, but is preserved in the topologically trivial ones. At the neutrality point ($\nu = 0$) all the bands are trivial. In all insulating states the Chern numbers of the occupied flat bands add up to 1 for odd filling factors and zero otherwise.

We also checked the nature of the ground state at non-integer filling factors according to the unrestricted HF calculation. In Fig. 5 we plot the band structure for $\varepsilon = 6$ and $\nu = 0.5$. The Fermi surface has two sheets, both with the same spin ($-\sigma$). In all cases, we find that the chemical potential crosses the low energy bands and gives rise to a spin polarized metal.

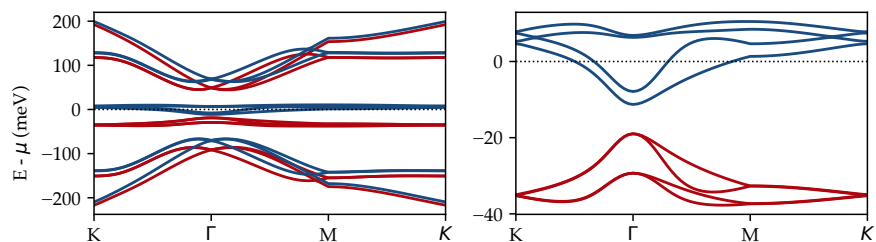

FIG. 5. Band structure of the low energy bands at $\nu = 0.5$ for $\varepsilon = 6$ showing some of the remote bands. Red lines: spin $\sigma$; Blue lines: spin $-\sigma$. On the right, we zoom in around the low energy flat bands. The chemical potential $\mu$ crosses the spin $-\sigma$ bands (blue lines) above the many-body gap. The ground state is described by a spin polarized metal.



\begin{thebibliography}{10}
\bibitem{note2} In TBG, different valleys are coupled by hopping between the monolayers. At the tight-binding level,
each state in the moire BZ is a non-trivial momentum dependent mixture of the two valleys. See Ref. \cite{Ramires}.
\bibitem{note3} Previous attempts to incorporate remote band effects
were mostly based on self-energy corrections to the flat bands in
the form of interaction assisted hopping terms. See Ref. \cite{Potasz,Liao,Kang3,Guinea}.
\bibitem{note4} Particle-hole symmetry is defined as invariance under the 
transformation $\nu\to -\nu$. See Ref. \cite{Potasz}.
\bibitem{Note} We set $t(\mathbf{r})=\cos^{2}\theta_{z}V_{\sigma}(\mathbf{r})+\sin^{2}\theta_{z}V_{\pi}(\mathbf{r})$,
where $\cos\theta_{z}=d/|\mathbf{r}|$, with $d$ the vertical
distance between sites. $V_{\sigma}(\mathbf{r})=V_{\sigma}^{0}\text{exp}\left[-(|\mathbf{r}|-d_{0})/\lambda\right]$ and 
 $V_{\pi}(\mathbf{r})=V_{\pi}^{0}\text{exp}\left[-(|\mathbf{r}|-a/\sqrt{3})/\lambda\right]$
are Slater-Koster functions, 
with $V_{\sigma}^{0}=-0.48$eV, $V_{\pi}^{0}=0.27$eV, and $\lambda=0.4525\text{Å}$.
$d_{0}=3.35\text{Å}$ is the unrelaxed interlayer distance and $a=2.42\text{Å}$
the in-plane lattice constant. See Ref. \cite{Nam}.
\end{thebibliography}
\end{document}